\documentclass{aa}
\usepackage{graphicx,natbib,amssymb}
\bibpunct{(}{)}{;}{a}{}{,}
\usepackage{txfonts}
\begin{document}
   \title{Geometry of giant star model atmospheres: A consistency test}

   \titlerunning{Geometry test}

   \author{U. Heiter
          \and
          K. Eriksson
          }

   \offprints{U. Heiter}

   \institute{Department of Astronomy and Space Physics,
              Uppsala University, Box 515, SE-75120 Uppsala\\
              \email{Ulrike.Heiter@astro.uu.se, Kjell.Eriksson@astro.uu.se}
              }

   \date{Received ; accepted }

   \abstract
   {}
   {We investigate the effect of a geometric inconsistency in the calculation of synthetic spectra of giant stars.}
   {Spectra computed with model atmospheres calculated in spherical geometry while using the plane-parallel approximation for line formation calculations ($s\_p$), as well as the fully plane-parallel case ($p\_p$), are compared to the consistently spherical case ($s\_s$).}
   {We present abundance differences for solar metallicity models with $T_{\rm eff}$ ranging from 4000 to 6500~K and $\log g$ from 0.5 to 3.0 [cgs]. The effects are smaller for $s\_p$ calculations ($-0.1$~dex in the worst case) than for the $p\_p$ case (up to $+0.35$~dex for minority species and at most $-0.04$~dex for majority species), both with respect to the $s\_s$ case. In the $s\_p$ case the differences increase slightly with temperature, while in the $p\_p$ case they show a more complex behaviour. In both cases the effects decrease with increasing $\log g$ and increase with equivalent width.}
   {Within the parameter range of F, G and K giants, consistency seems to be less important than using a spherical model atmosphere. The abundance differences due to sphericity effects presented here can be used for error estimation in abundance studies relying on plane-parallel modelling.}
   

   \maketitle
%

\section{Introduction} \label{introduction}

Classical abundance analyses are usually based on model atmospheres calculated assuming plane-parallel geometry. This means that radiative transfer is solved in only one depth variable, neglecting the curvature of the atmosphere. For most stars this is justified because the extension of the atmosphere is negligibly small compared to the stellar radius. Extreme cases are cool giants and supergiants, where geometry effects due to the thickness of the atmosphere have to be taken into account. They have been studied accordingly well in the literature.

Sphericity effects on structures, low resolution spectra and colors of cool ($T_{\rm eff} \le 4000$~K) giants and supergiants have been studied for example by \citet{Plez:92}. They showed that for one of their hottest models with ($T_{\rm eff}$,$\log g$\footnote{$\log g$ values are given in cgs units throughout the paper.},mass)=(3800~K,1.0,1~$M_{\odot}$), the changes in temperature structure when releasing the plane-parallel approximation are relatively small (about 35~K at $\log \tau_{Ross}=-4$, the atmospheric extension being about 4\%).
\citet{Plez:90} discussed effects of atmospheric extension on the determination of effective temperature, surface gravity and abundances of cool supergiants ($T_{\rm eff} \le 4500$~K). He noted that the effects are not linear with temperature and gravity.

Spherical model atmospheres for a large range of giant star temperatures have been presented by \citet{Haus:99}. They also presented changes in spectral flux with respect to plane-parallel models. For a model with ($T_{\rm eff}$,$\log g$)=(5600~K,0.0) they show for example that the change in central depth of individual spectral lines can be up to 15\%.

F to K giants are commonly used to study abundances of stellar systems, mainly due to their intrinsic brightness. With respect to atmosphere geometry they represent borderline cases and are therefore usually analysed using plane-parallel radiative transfer.
Many examples for recent large scale abundance analyses of giant stars can be found in the literature. Cepheid stars at various pulsation phases were studied by \citet{Luck:04}. These stars are used to determine the Galactic abundance gradient \citep{Andr:04}. Another example is the analysis of probable giant star planet hosts \citep{Take:05}.
Furthermore, red giants are frequently contained in samples of Galactic halo stars, which are used to study Galactic chemical evolution and nucleosynthesis by means of classical abundance analysis.
Earlier examples are the large scale high resolution abundance studies of metal poor stars by \citet[][31 stars with $\log g \lesssim 2.5$]{McWi:95} and \citet[][7 stars with $\log g \lesssim 2.0$]{Ryan:96}, whose results have been used by \citet{Trav:01} in their study of the chemical evolution of the Galactic halo.
More recently, \citet{Simm:04} determined abundances of Fe, La and Eu for stars over a wide range of metallicities. A substantial fraction of those are giant stars in the $\log g$ range of 0.5 to 2.0, all with [Fe/H] $\lesssim -1$~dex.

In recent years, grids of model atmospheres calculated in spherical geometry have become available to an increasing extent \citep[e.g.][]{Gust:03,Haus:99}. However, line formation will probably continue to be computed in plane-parallel geometry in abundance analysis procedures. In this paper, we investigate the effect of such an inconsistency on synthetic spectra and abundances of giant stars. We then compare to the effects due to using a consistent, but plane-parallel combination of model atmosphere and spectrum synthesis.

\begin{figure}
   \resizebox{\hsize}{!}{\includegraphics[]{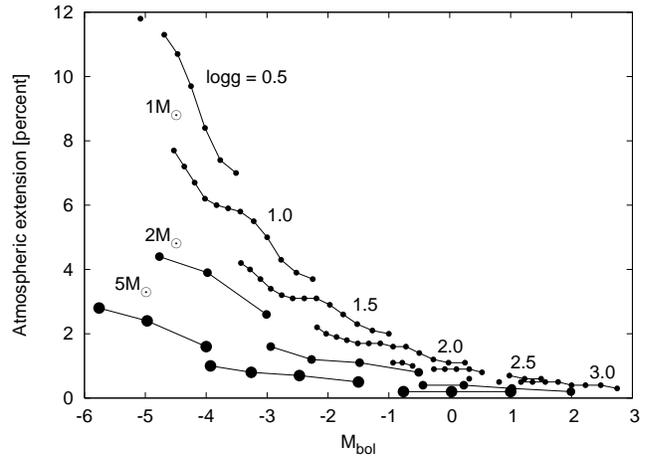}}
   \caption{Atmospheric extension ($\frac{r(\tau_{Ross}=10^{-5})}{r(\tau_{Ross}=1)}-1$) of spherical models with $\xi_{\rm t}$ = 2~km~s$^{-1}$ and solar abundances as a function of bolometric magnitude for $M$ = 1, 2, and 5~$M_{\odot}$ as indicated by increasing symbol size.
            Models with equal $\log g$ are joined by lines, except where certain $T_{\rm eff}$ values are missing, and are labelled with their $\log g$ value.
            Model temperatures range from 4000 to 7000~K in steps of 250~K for $M$ = 1~$M_{\odot}$ and 1000~K for $M$ = 2 and 5~$M_{\odot}$, increasing from right to left along lines.
           }
   \label{extension}
   \end{figure}

Figure~\ref{extension} shows the atmospheric extension of spherical MARCS models (see Section~\ref{models}) with $\xi_{\rm t}$ = 2~km~s$^{-1}$, solar abundances and various temperatures and gravities as a function of bolometric magnitude.
Here, the extension is defined as the difference in radius at the points in the atmosphere where $\tau_{Ross}=10^{-5}$ and 1 relative to the radius at $\tau_{Ross}=1$.
The bolometric magnitude has been calculated from the model fluxes on the scale defined by IAU Commission 36 \citep{Ande:99}.
This figure can be directly compared to Fig.~1 of \citet{Plez:92}. It shows that atmospheric extensions increase relatively rapidly with effective temperature for $\log g < 2$. One can also see small local maxima at about 5250~K, similar to the ones at 3200~K found by \citet{Plez:92}. These features in the shape of lines of equal $\log g$ can probably be attributed to a decrease in partial pressure gradient of diatomic molecules such as CO or CN.
The same plot for models with abundances corresponding to halo stars with [Fe/H] $= -2$~dex looks very similar, except that the aforementioned maxima are not present. This is in accordance with the low metallicity where molecules are not as important.

\begin{figure}
   \resizebox{\hsize}{!}{\includegraphics[]{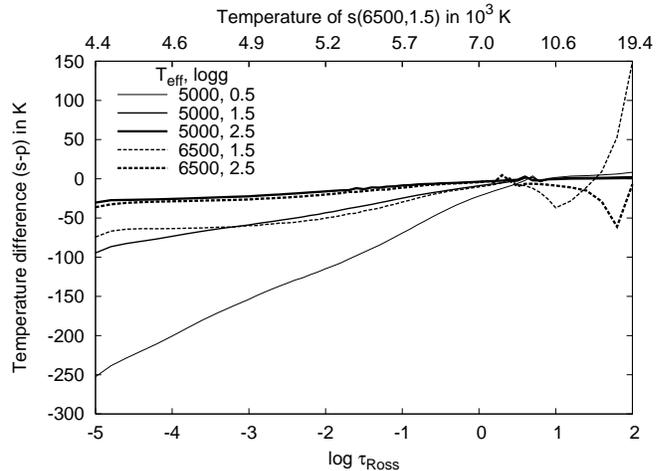}}
   \caption{Differences between spherical and plane-parallel model temperatures ($T_s-T_p$) as a function of Rosseland optical depth of models with $\xi_{\rm t}$ = 2~km~s$^{-1}$, solar abundances and $M = 1~M_{\odot}$. Solid (dashed) lines show $T_{\rm eff}$ = 5000~K (6500~K) models and line width increases with $\log g$. The scale on the top horizontal axis gives the temperature of the spherical ($T_{\rm eff}$,$\log g$)=(6500~K, 1.5) model in units of $10^3$~K.}
   \label{moddiff}
   \end{figure}

Figure~\ref{moddiff} shows how the temperature structures of spherical models with solar abundances in the parameter range investigated here differ from plane-parallel ones. The dilution of the radiation field in spherical geometry causes the spherical model to be cooler than the plane-parallel one above an optical depth of $\approx$ 1, and the temperature difference increases outwards.

The differences seen for the hotter models at large optical depths can be attributed to numerical effects in the linearized mixing length equations. For these models, the lower boundary of the convection zone is very close to the lower boundary of the atmosphere. Note that these differences are not visible in the spectra and that the differences are less than 1\% (see temperature scale for the spherical ($T_{\rm eff}$,$\log g$)=(6500~K, 1.5) model at the top axis of the figure.

In the outermost layers, the temperature differences for models of equal $\log g$ and different $T_{\rm eff}$ show a complex behaviour. In certain cases, models with larger extension show less surface cooling with respect to the plane-parallel counterpart than less extended models, for example the models with $\log g$=1.5 shown in Figure~\ref{moddiff}. This is very likely due to the effect of the change in temperature structure on molecular absorption. An inspection of the grid models shows that the changes in partial pressures of molecules in the outermost layers, when going from plane-parallel to spherical geometry, do not simply scale with effective temperature. This applies in particular to molecules containing carbon, as well as TiO and SiS. Since molecular absorption can either cool or heat the outer layers \citep{Gust:79}, the net effect might well be more cooling in a less extended model as compared to a more extended model.

\section{Models, spectra and equivalent widths} \label{models}

We calculated spectra with two different versions of the spectrum synthesis code used in Uppsala (plane-parallel -- $p$ -- and spherical -- $s$) for representative model parameters.
We chose solar metallicity for all models and masses of 1, 2 and 5~$M_{\odot}$ for $s$ models.
The microturbulence parameter $\xi_{\rm t}$ was set to 2 and 5~km~s$^{-1}$.
$T_{\rm eff}$ was varied between 4000 and 6500~K in steps of 500~K for 1~$M_{\odot}$ and 1000~K for 2 and 5~$M_{\odot}$.
The $\log g$ values ranged from 0.5 to 3.0 in steps of 0.5 for 1~$M_{\odot}$ and were set to 1.0, 2.0 and 3.0 for 2 and 5~$M_{\odot}$.
For 1~$M_{\odot}$, this corresponds to radii at $\tau_{Ross}=1$ of $\approx5R_{\odot}$ ($\log g$ = 3.0), $\approx15R_{\odot}$ ($\log g$ = 2.0) and $\approx90R_{\odot}$ ($\log g$ = 0.5); for 2 and 5~$M_{\odot}$ and $\log g$ = 2.0, the radii are $\approx20$ and $40R_{\odot}$, respectively.
For $\log g$ = 0.5, only models with $T_{\rm eff} \le$ 5000~K were available.
Normalized synthetic spectra were calculated for wavelengths between 5400 and 7200~\AA\ using atomic line data from VALD \citep{Kupk:99} and MARCS model structures \citep{Gust:03}\footnote{http://marcs.astro.uu.se} as input. For a discussion of the method for solving the radiative transfer equations in the atmospheric models and spectrum synthesis see \citet{Plez:92} and \citet{Nord:84}. We regard three different combinations of model atmosphere ({\em atm}) and spectrum synthesis ({\em syn}) geometries: {\em atm\_syn} = {\em s\_s} (consistently spherical), {\em s\_p} (inconsistent), {\em p\_p} (consistently plane-parallel).

As a next step, we selected all lines in this spectral region which were found to be not blended by more than 30~\% by neighboring lines. This line selection was done for the (5000~K,0.5), $\xi_{\rm t}=$2 and 5~km~s$^{-1}$, 1~$M_{\odot}$ $s\_s$ models. Equivalent widths were then calculated for these lines for the $s\_s$ models with an equivalent width / abundance fit version of the spectrum synthesis code. Finally, abundance differences were obtained for $s\_p$ and $p\_p$ models by fitting the calculated equivalent widths to the $s\_s$ ones. Table~\ref{species} lists all species contained in the line list and the number of lines selected for each of them, as well as the fraction of the element in the given ionization stage, averaged over $-1.0 \le \log\tau_{Ross} \le -0.1$.

   \begin{table}
      \caption[]{Number of lines selected for each species in the geometrically consistent ($s\_s$) synthetic spectrum with ($T_{\rm eff}$,$\log g$,$\xi_{\rm t}$)=(5000~K,0.5,2~km~s$^{-1}$) with equivalent widths between 5 and 250~m\AA, as well as fraction of the element in the given ionization stage (in \%, averaged over $-1.0 \le \log\tau_{Ross} \le -0.1$).}
         \label{species}
     $$ 
         \begin{array}{p{8mm}rrp{2mm}p{8mm}rrp{2mm}p{8mm}rrp{2mm}p{8mm}rr}
            \hline
            \noalign{\smallskip}
Species &  $n$ & \% & & Species & $n$ & \% & & Species & $n$ & \% \\
            \noalign{\smallskip}
            \hline
            \noalign{\smallskip}
Al I  &   9 &   0.1 & & Gd II &  14 &  99.8 & & Sc II &  15 &  99.9  \\
C  I  &  21 & 100.0 & & La II &  33 &  98.6 & & Si I  & 173 &   5.2  \\
Ca I  &  28 &  0.01 & & Mg I  &   8 &   0.5 & & Si II &   4 &  94.8  \\
Ce II &  33 &  98.3 & & Mn I  &  17 &   0.5 & & Sm II & 104 &  98.3  \\
Co I  &  76 &   1.8 & & Mn II &   3 &  99.5 & & Ti I  & 115 &   0.1  \\
Cr I  &  41 &   0.1 & & Mo I  &   3 &   0.3 & & Ti II &  34 &  99.9  \\
Cr II &  22 &  99.9 & & Nd II & 109 &  96.7 & & V  I  &  57 &   0.1  \\
Dy II &   5 &  99.3 & & Ni I  & 124 &   2.6 & & V  II &  21 &  99.9  \\
Er II &   5 &  99.6 & & O  I  &   4 & 100.0 & & Y  II &  10 &  99.8  \\
Eu II &   7 &  98.6 & & Pr II &  73 &  94.8 & & Zr I  &   6 &   0.1  \\
Fe I  & 562 &   1.0 & & S  I  &  21 &  91.1 & & Zr II &   7 &  99.9  \\
Fe II &  68 &  99.0 & & Sc I  &   8 &  0.03 & &       &     &        \\
            \noalign{\smallskip}
            \hline
         \end{array}
     $$ 
   \end{table}
%
   \begin{figure*}
   \sidecaption
   \includegraphics[width=12cm,angle=90]{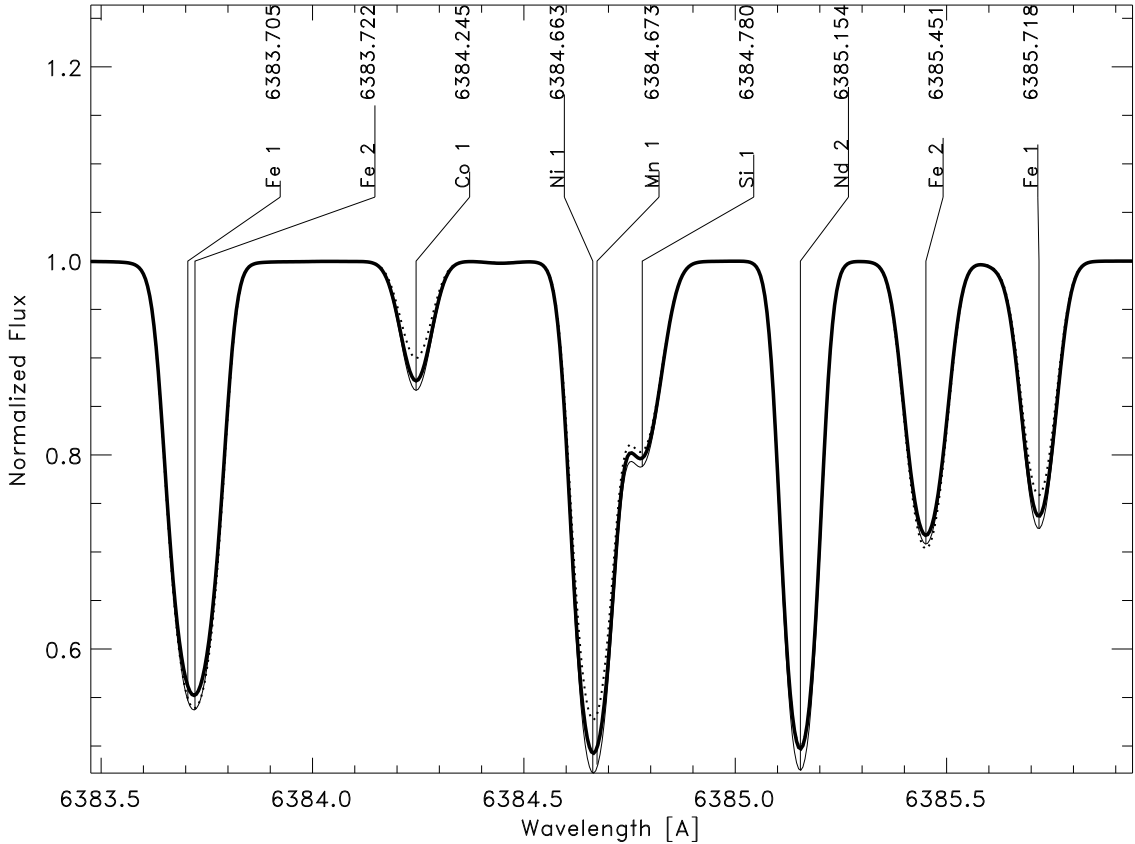}
   \caption{Comparison of $s\_s$ (thick line), $s\_p$ (thin line) and $p\_p$ (dotted line) spectra for ($T_{\rm eff}$,$\log g$,$\xi_{\rm t}$,$M$)=(5000~K,0.5,2~km~s$^{-1}$,1~$M_{\odot}$).}
   \label{profiles}
   \end{figure*}
%

%
   \begin{figure*}
   \sidecaption
   \includegraphics[width=12cm]{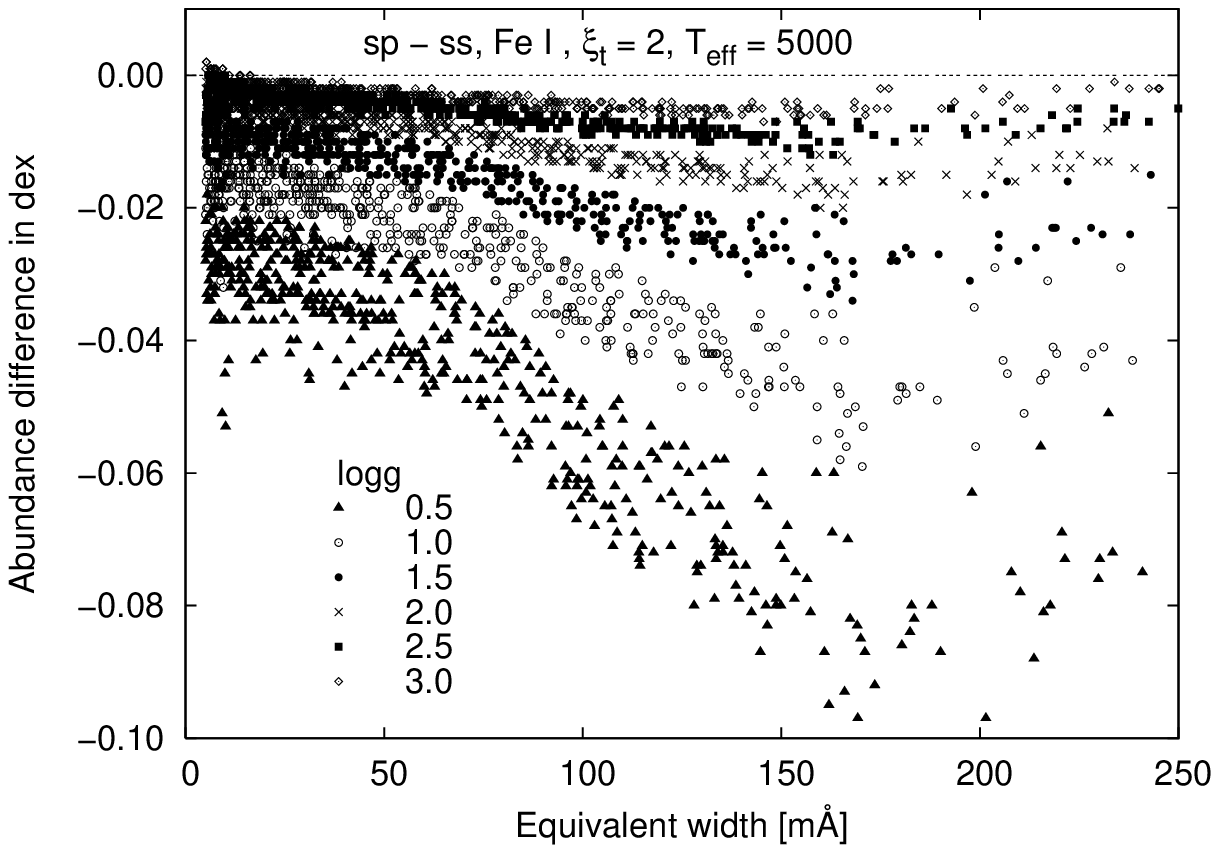}\\
   \includegraphics[width=12cm]{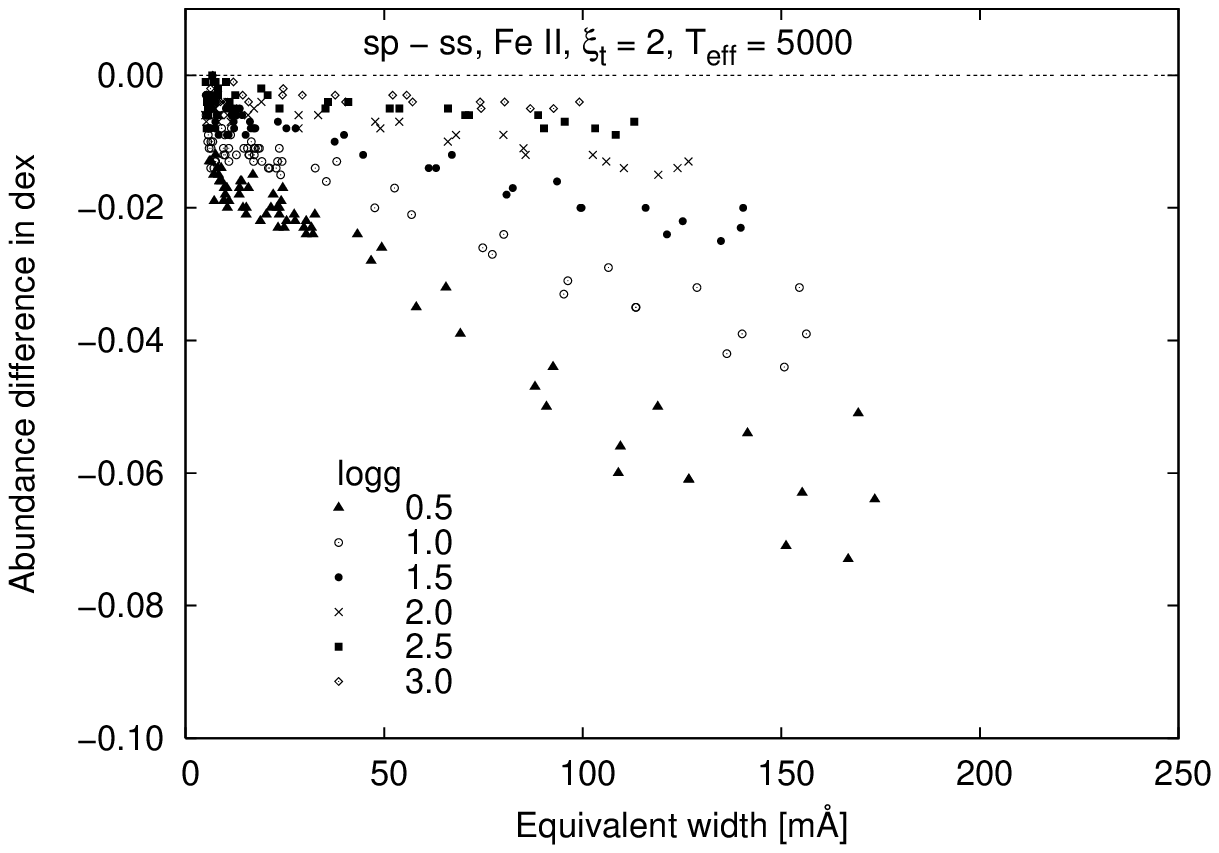}
   \caption{Abundance differences for inconsistent and consistently spherical equivalent widths of Fe~I lines (upper plot) and Fe~II lines (lower plot) as a function of $s\_s$ equivalent widths for models with a mass of 1~$M_{\odot}$ with $\xi_{\rm t}=$2~km~s$^{-1}$, $T_{\rm eff}$=5000~K, and different $\log g$ values as indicated by different symbols.}
   \label{abunddiff_s_Fe_logg}
   \end{figure*}
%

%
   \begin{figure*}
   \sidecaption
   \includegraphics[width=12cm]{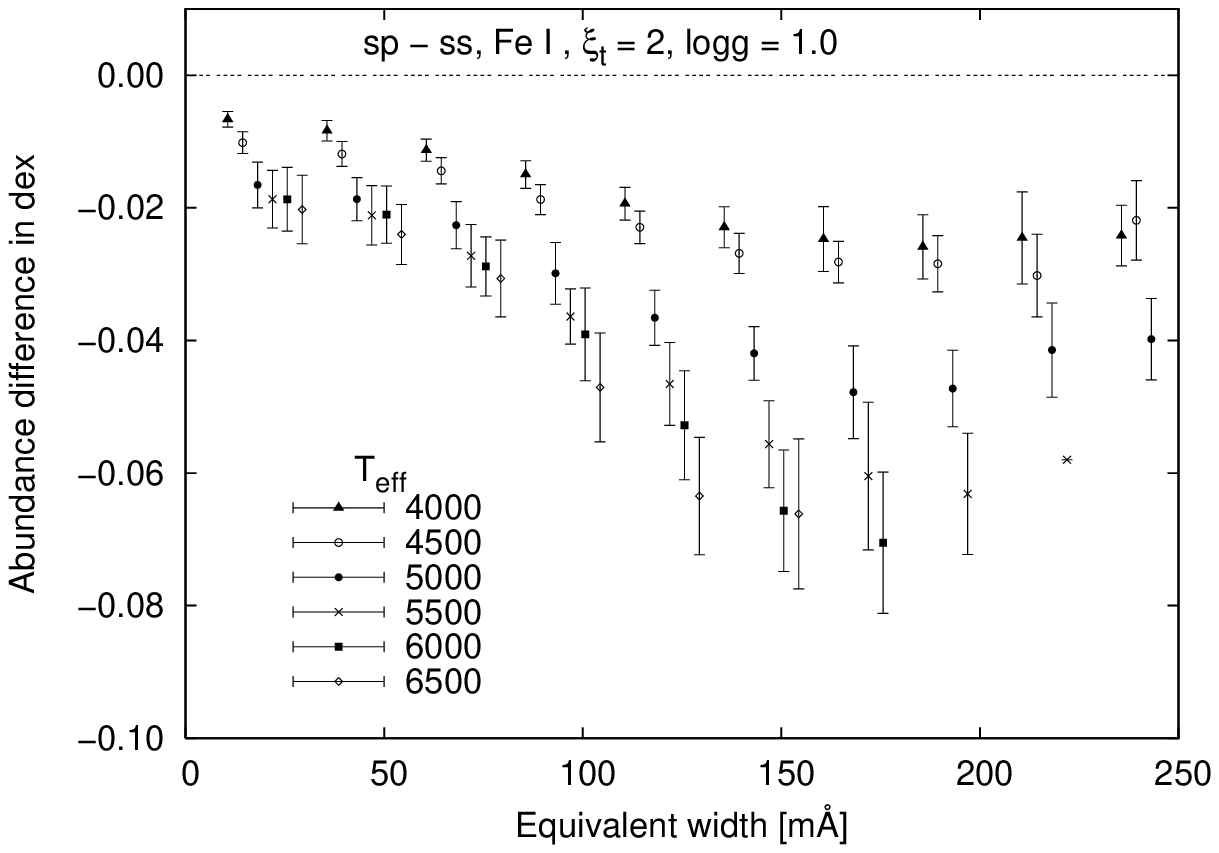}\\
   \includegraphics[width=12cm]{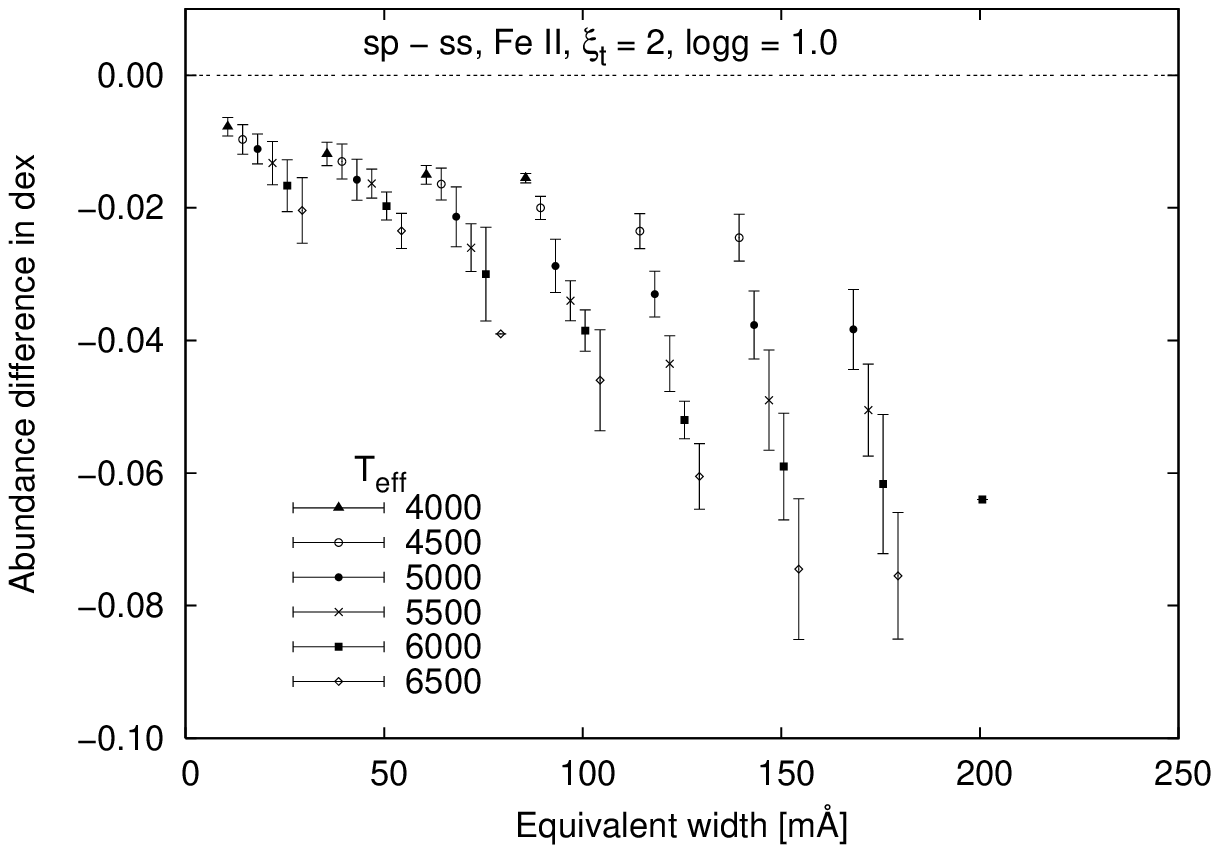}
   \caption{Abundance differences for inconsistent and consistently spherical equivalent widths of Fe~I lines (upper plot) and Fe~II lines (lower plot) as a function of $s\_s$ equivalent widths for models with a mass of 1~$M_{\odot}$ with $\xi_{\rm t}=$2~km~s$^{-1}$, $\log g$=1.0, and different $T_{\rm eff}$ values as indicated by different symbols. The differences have been averaged over bins of 25~m\AA, with bin centers offset for clarity and error bars showing the standard deviations.}
   \label{abunddiff_s_Fe_Teff}
   \end{figure*}
%

%
   \begin{figure*}
   \sidecaption
   \includegraphics[width=12cm]{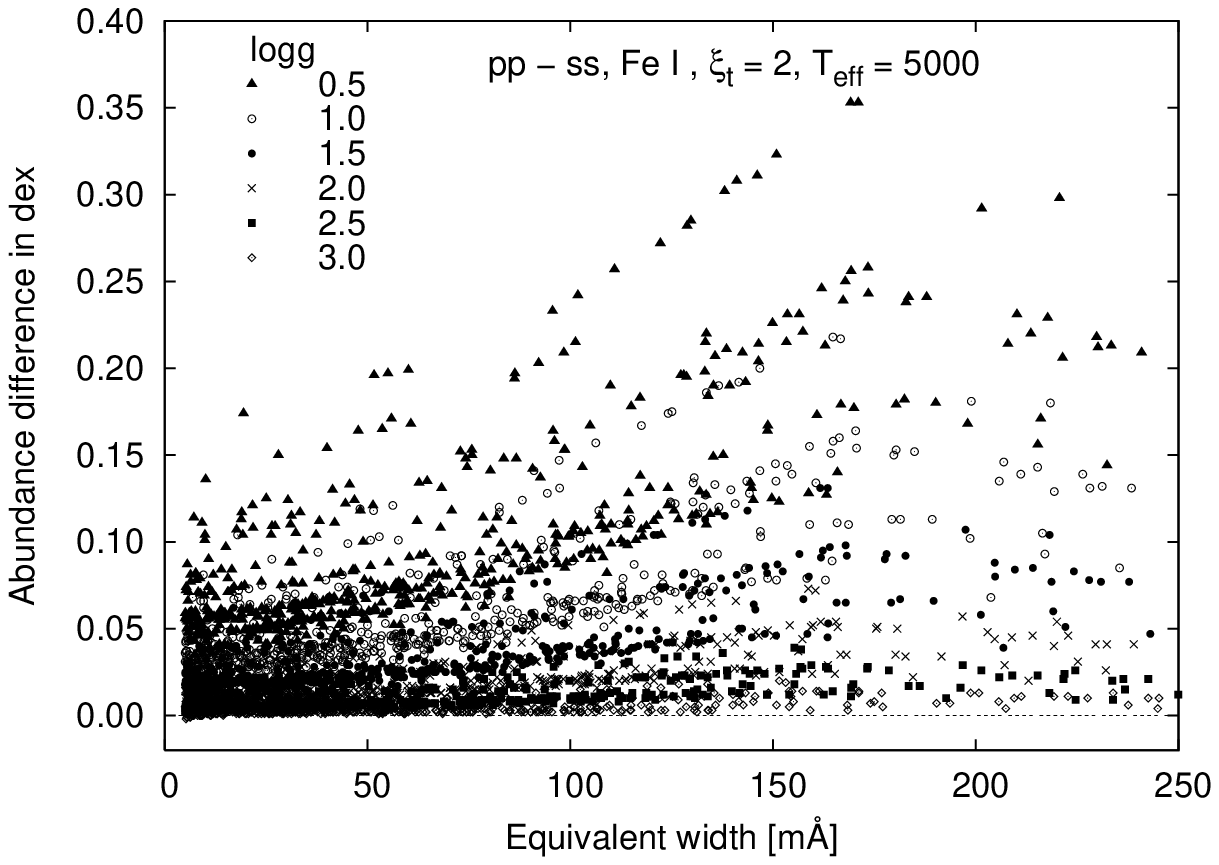}\\
   \includegraphics[width=12cm]{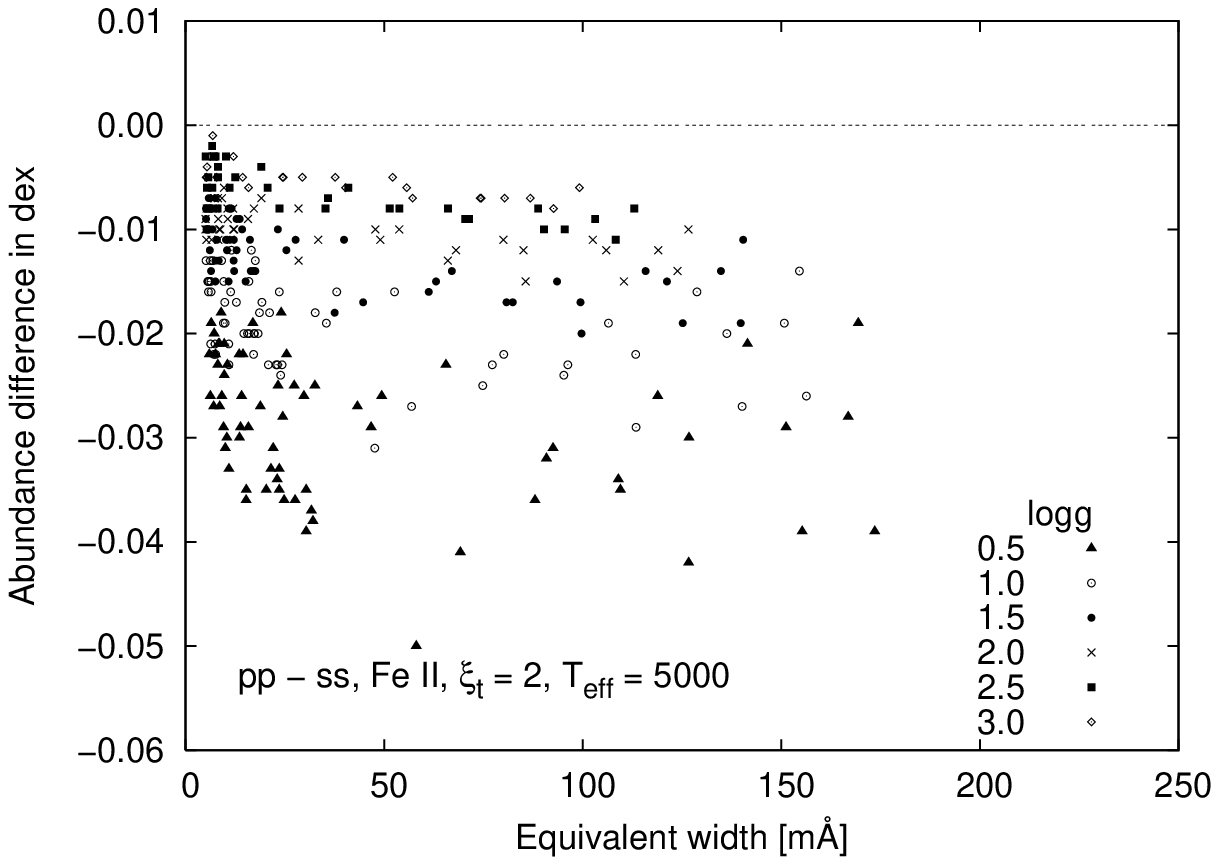}
   \caption{Differences between consistently plane-parallel and spherical abundances versus $s\_s$ equivalent widths of Fe~I lines (upper plot) and Fe~II lines (lower plot) for models with a mass of 1~$M_{\odot}$ with $\xi_{\rm t}=$2~km~s$^{-1}$, $T_{\rm eff}$=5000~K, and different $\log g$ values as indicated by different symbols.}
   \label{abunddiff_p_Fe_logg}
   \end{figure*}
%

%
   \begin{figure*}
   \sidecaption
   \includegraphics[width=12cm]{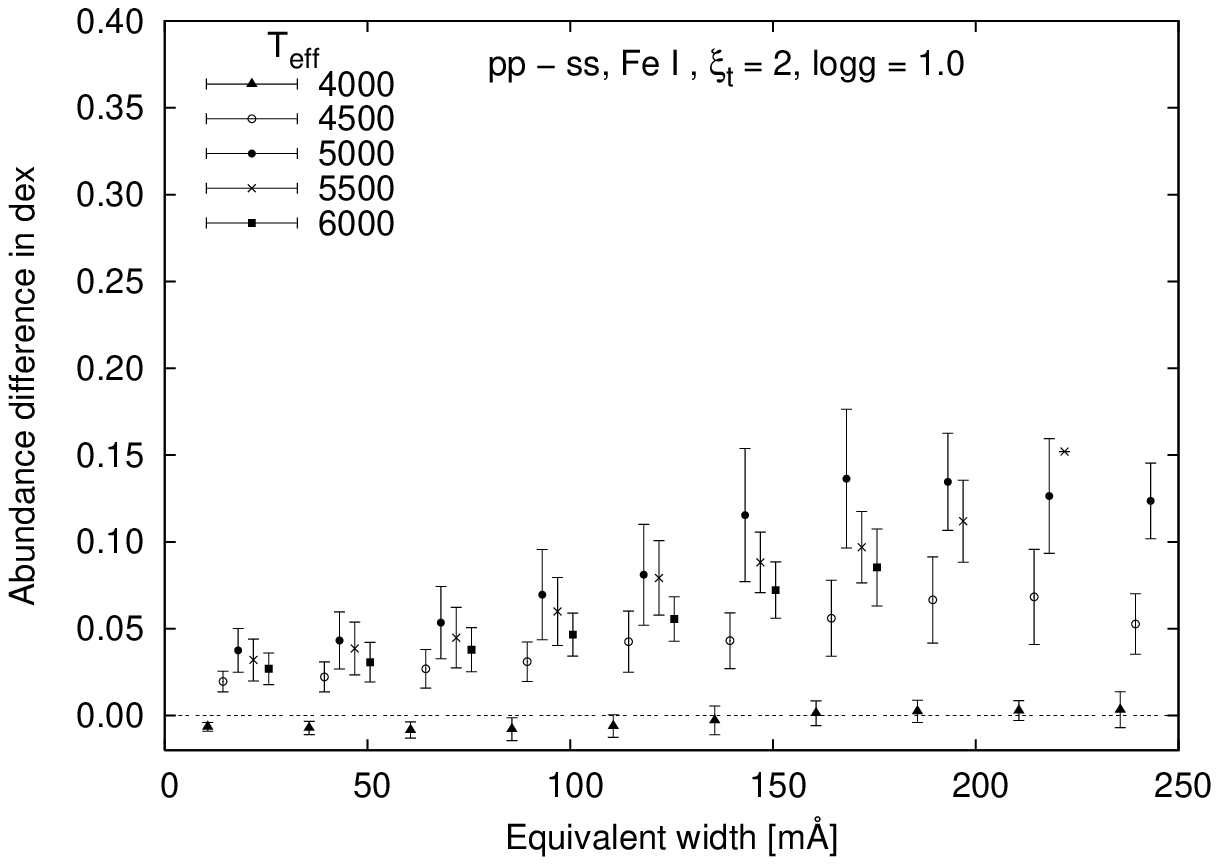}\\
   \includegraphics[width=12cm]{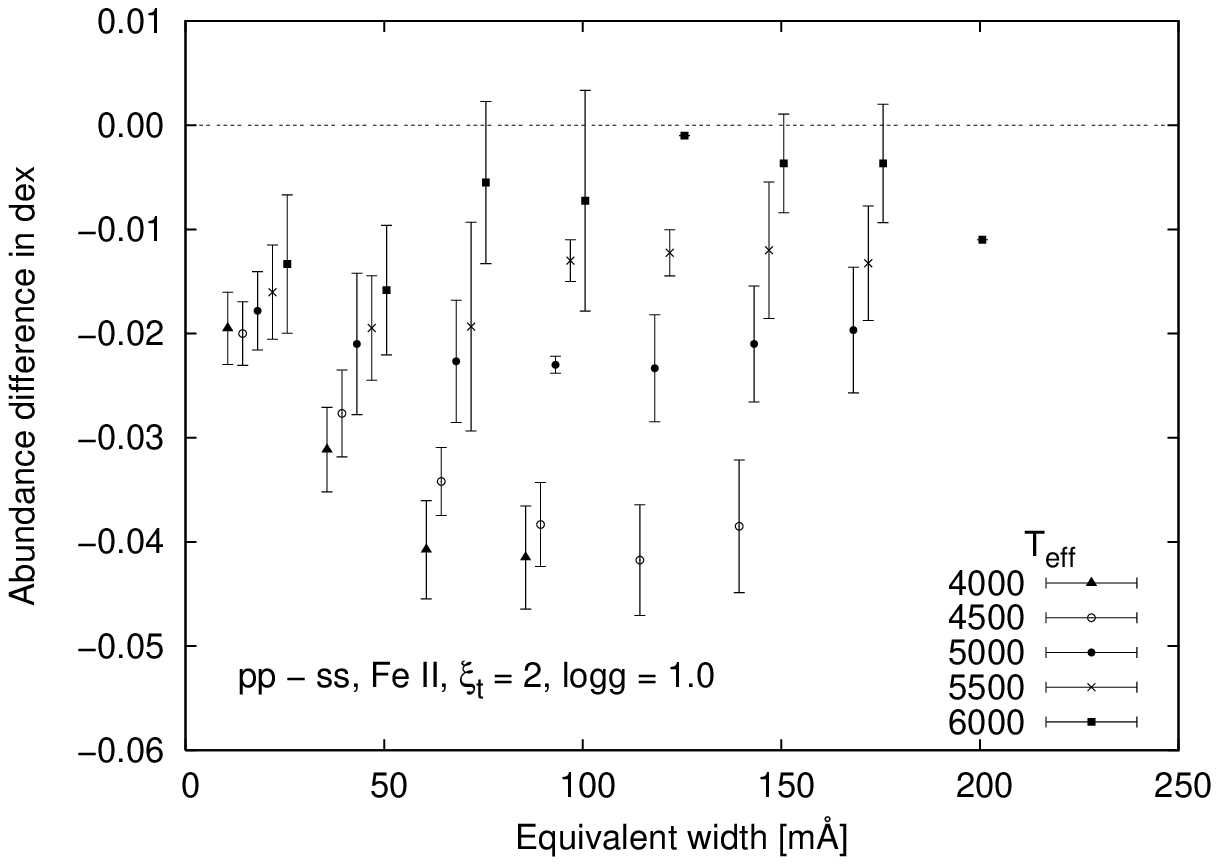}
   \caption{Differences between consistently plane-parallel and spherical abundances versus $s\_s$ equivalent widths of Fe~I lines (upper plot) and Fe~II lines (lower plot) for models with a mass of 1~$M_{\odot}$ with $\xi_{\rm t}=$2~km~s$^{-1}$, $\log g$=1.0, and different $T_{\rm eff}$ values as indicated by different symbols. The differences have been averaged over bins of 25~m\AA, with bin centers offset for clarity and error bars showing the standard deviations.}
   \label{abunddiff_p_Fe_Teff}
   \end{figure*}
%

\section{Results} \label{results}

We take the consistent $s\_s$ spectra as a reference and present a comparison to the spectra with the two other geometry combinations. Fig.~\ref{profiles} illustrates the geometry effects on line profiles. One can see that mainly the line centers are affected, which can be ascribed to the way the model atmosphere structure changes (see Fig.~\ref{moddiff}). The following two subsections summarize the effects on line abundances for 1~$M_{\odot}$ models. We restrict the investigation to lines with equivalent widths between 5 and 250~m\AA.

\subsection{Effects of inconsistent spectrum synthesis} \label{s}

Figure~\ref{abunddiff_s_Fe_logg} shows the differences between $s\_p$ and $s\_s$ abundances versus $s\_s$ equivalent widths ($W$) for Fe~I lines (upper plot) and Fe~II lines (lower plot). Models with $T_{\rm eff}$ = 5000~K, low microturbulence and various gravities are shown. Apart from the expected increase towards lower gravities, one notices that the effect is on average about 0.01~dex smaller for Fe~II than for Fe~I, for lines with $W \lesssim 50$~m\AA\ and $\log g \le 1$. The same difference between ionization stages is seen for all other elements with at least 10 lines of each of the two ionization stages (Cr, Ti, V).
The differences are slightly larger for higher $T_{\rm eff}$s. This is illustrated in Fig.~\ref{abunddiff_s_Fe_Teff}, which shows the differences for Fe lines for $\xi_{\rm t}$ = 2~km~s$^{-1}$, $\log g = 1.0$ and several $T_{\rm eff}$ values, averaged over bins of 25~m\AA. The trends are weaker for $\xi_{\rm t}$ = 5~km~s$^{-1}$ and different for different species.
The differences are in general larger for the lower $\xi_{\rm t}$ value, where line saturation sets in at smaller $W$.
Neutral species show the same differences as Fe~I, except for C~I, S~I and Si~I, for which the effect is on average smaller by 0.01~dex for the lowest $\log g$ values and weak lines. 
Singly ionized species show the same differences as Fe~II, except for weak lines of rare earth elements for $\log g = 0.5$. For Ce~II, the effect is on average smaller by 0.01~dex, and for Gd~II, Pr~II, Nd~II, La~II, Sm~II, the effect is on average larger by 0.01~dex.

\subsection{Effects of consistent plane-parallel spectrum synthesis} \label{p}

Figure~\ref{abunddiff_p_Fe_logg} shows the differences between $p\_p$ and $s\_s$ abundances versus $s\_s$ equivalent widths for Fe~I and Fe~II lines, for the models with $T_{\rm eff}$ = 5000~K, low microturbulence and various gravities. In this case, neutral and ionized Fe lines show an opposite effect, which is also seen for all other elements.
The dependence of abundance differences on $T_{\rm eff}$ is more complex than in the previous case (Section~\ref{s}), as shown Figure~\ref{abunddiff_s_Fe_logg}. For Fe~I lines, the differences are essentially zero at $T_{\rm eff}$ = 4000~K, increase up to $T_{\rm eff}$ = 5000~K, and decrease slightly towards higher $T_{\rm eff}$. For Fe~II lines, the differences decrease with increasing $T_{\rm eff}$ over the whole range.

Again, most neutral species behave similarly to Fe~I, with the following exceptions:
\begin{itemize}
   \item C~I, O~I and S~I (all predominantly neutral): The effect has opposite sign to that of Fe~I (similar to Fe~II).
   \item Si~I: The effect is less than half that of Fe~I.
   \item Ti~I and V~I: The effect is on average larger than for Fe~I.
\end{itemize}

Most singly ionized species show the same differences as Fe~II. Exceptions are again the rare earth elements, but also Sc and Ti:
\begin{itemize}
   \item Ce~II, La~II, Nd~II, Sc~II: The effect is smaller than for Fe~II and changes sign for large equivalent widths.
   \item Sm~II, Ti~II: The effect is on average 0.01~dex smaller than for Fe~II for $\log g \le 1.0$.
\end{itemize}

   \begin{figure*}
   \sidecaption
   \includegraphics[width=12cm]{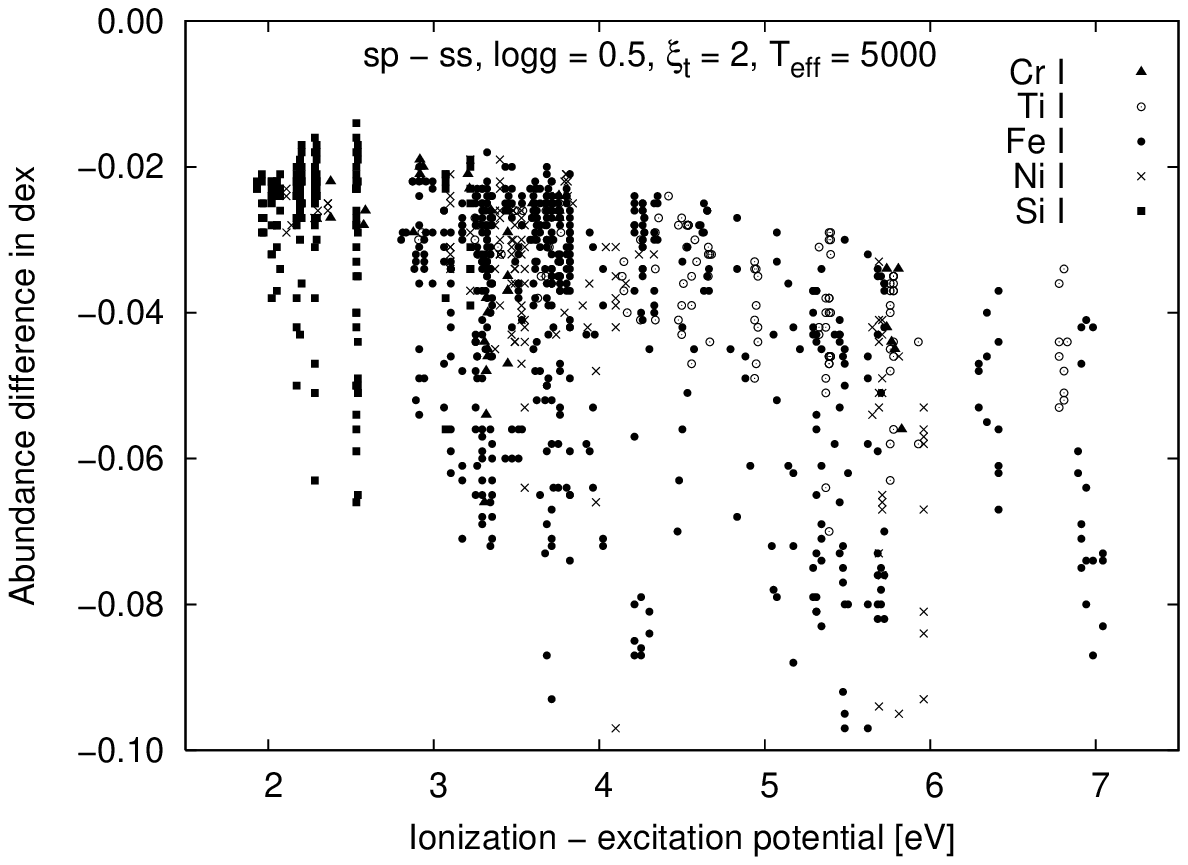}\\
   \includegraphics[width=12cm]{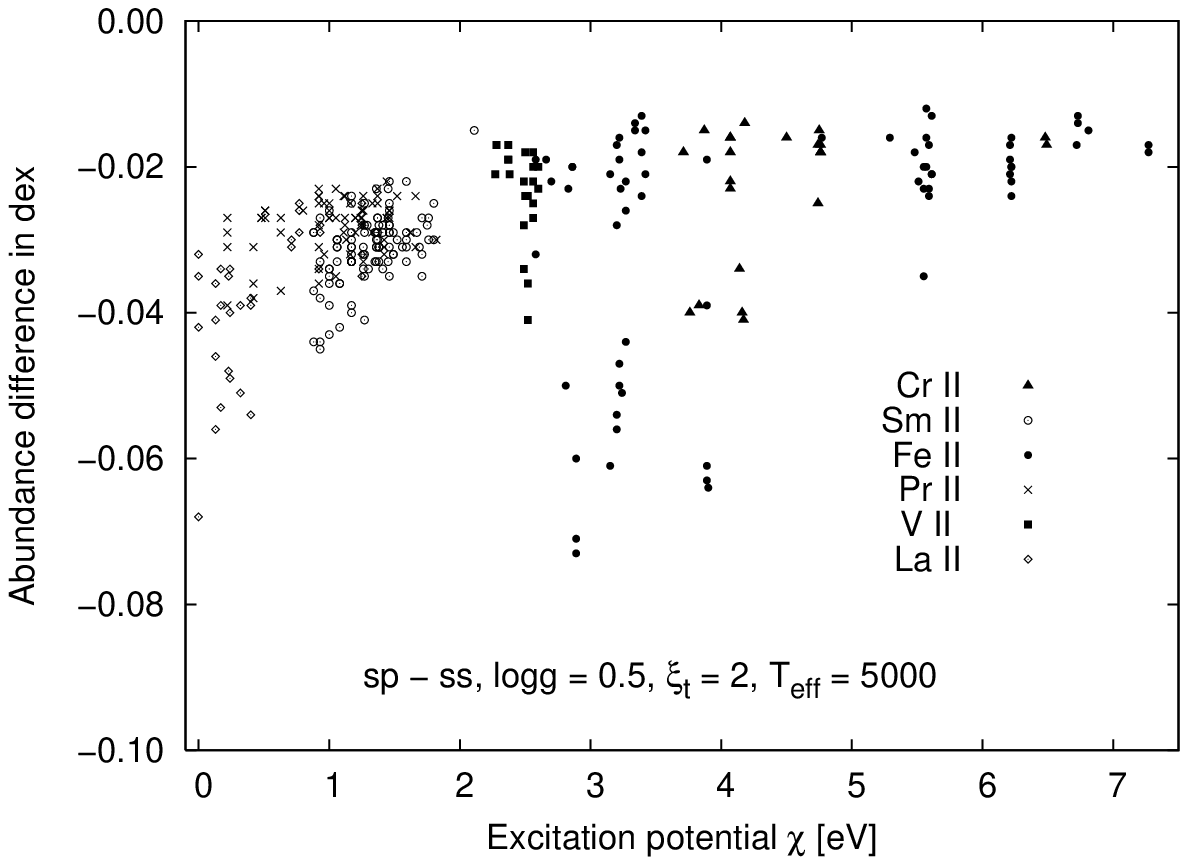}
   \caption{Abundance differences $s\_p - s\_s$ versus difference of ionization and excitation potentials (neutral species, upper plot) and versus excitation potential (singly ionized species, lower plot).}
   \label{exc_s}
   \end{figure*}

   \begin{figure*}
   \sidecaption
   \includegraphics[width=12cm]{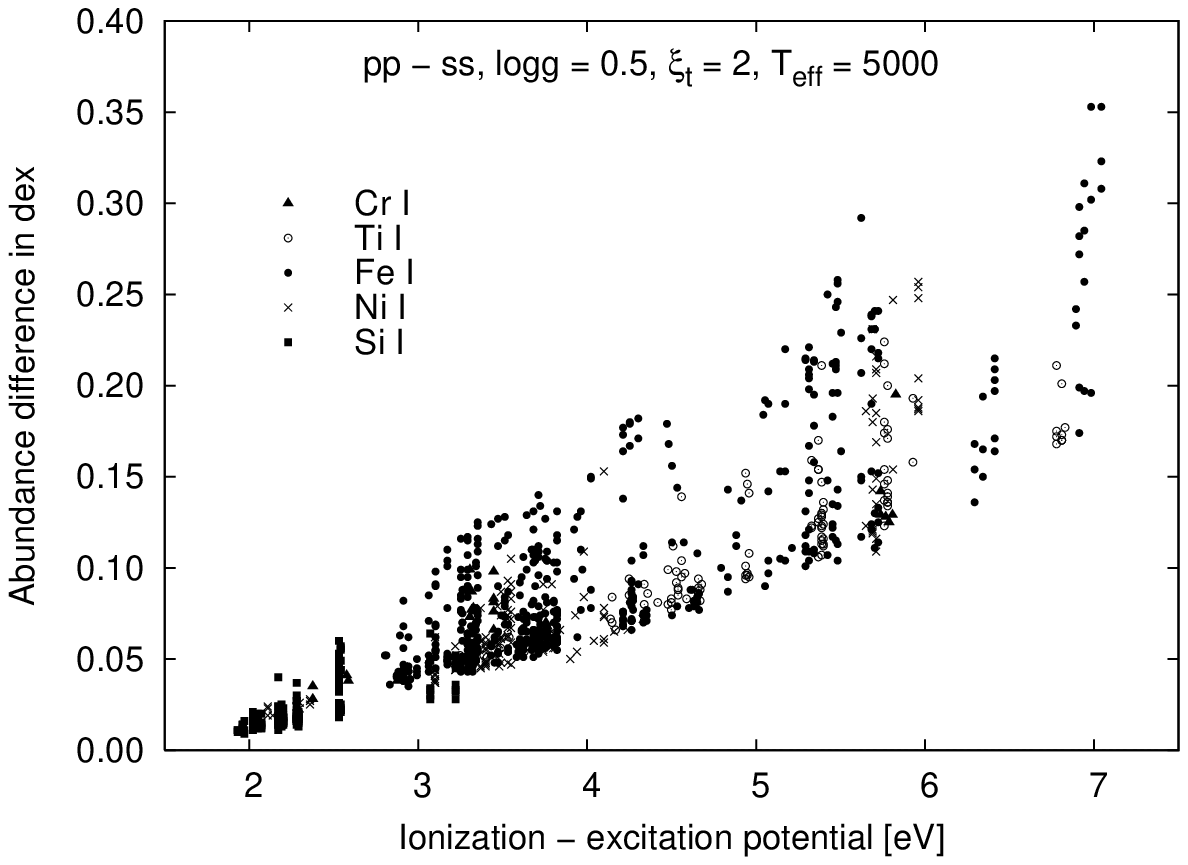}\\
   \includegraphics[width=12cm]{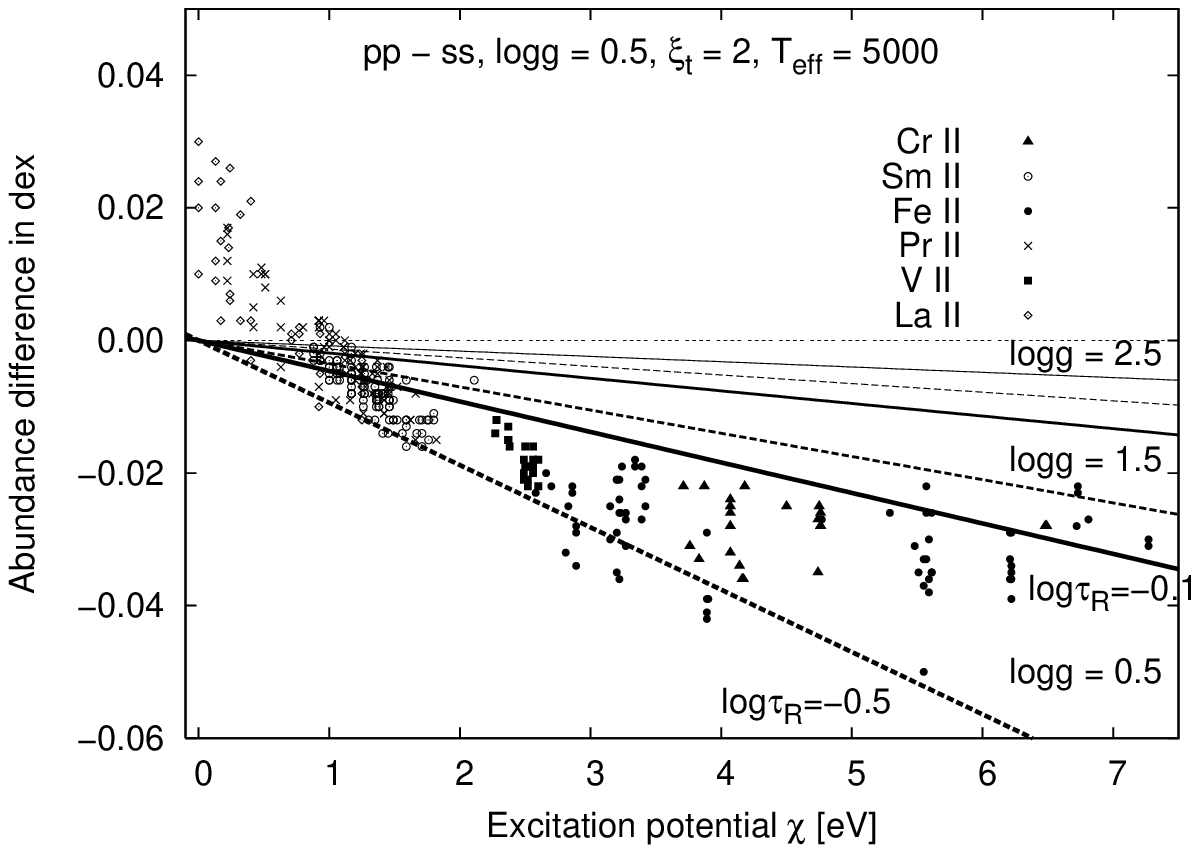}
   \caption{Abundance differences $p\_p - s\_s$ versus difference of ionization and excitation potentials (neutral species, upper plot) and versus excitation potential (singly ionized species, lower plot). The lines in the lower plot indicate the variation of abundance difference as estimated from the difference in model temperature only (see text, Section~\ref{analytic_exc_sect}) at $\log\tau_{\rm Ross}=-0.5$ (dashed) and $\log\tau_{\rm Ross}=-0.1$ (solid), for different $\log g$ values as indicated by different line width.}
   \label{exc_p}
   \end{figure*}

\subsection{Mass dependence}
For models with ($T_{\rm eff}$,$\log g$,$\xi_{\rm t}$)=(5000~K,1.0,2~km~s$^{-1}$) and a mass of 2~$M_{\odot}$, the abundance differences are smaller by 0.005 (weak lines) to 0.01~dex (strong lines) compared to 1~$M_{\odot}$ models. For a model mass of 5~$M_{\odot}$, they are smaller by 0.01 (weak lines and strong lines of majority species in the $p\_p$ case) to 0.03 (strong lines in the $s\_p$ case) to 0.05~dex (strong lines of minority species in the $p\_p$ case). In general, higher masses result in smaller effects and the dependence on mass is proportional to abundance difference. As can be seen in Fig.~\ref{extension}, the extensions in 5~$M_{\odot}$, $\log g$=1.0 models are approximately the same as in 1~$M_{\odot}$, $\log g$=1.5 models. We therefore expect that an increase in mass from 1 to 5~$M_{\odot}$ has the same effect as increasing $\log g$ by 0.5. This is confirmed by the calculations.

\section{Discussion}

\subsection{Trends of effects with line data}
It seems that the temperature structure used as input for the spectrum synthesis calculation is more important than the geometry used in such a code. Since the temperature differences decrease with depth in the atmosphere, spectral lines formed at different depths will be affected differently. We calculated average formation depths $\langle\tau_{\rm Ross}\rangle$ of the line centers for a few species which show different effects (Section~\ref{results}) for the $s$ model with ($T_{\rm eff}$,$\log g$,$\xi_{\rm t}$)=(5000~K,0.5,2~km~s$^{-1}$). The distributions (relative frequency of lines forming at different $\log\langle\tau_{\rm Ross}\rangle$) are rather similar, except for those of C~I, S~I, Si~I and Ti~I. They show a somewhat higher and narrower peak at $\log\langle\tau_{\rm Ross}\rangle \approx -0.3$ than that of Fe~I. This will introduce a small species to species difference in the geometry effects.

The largest part of the differential effect can however be attributed to the different distributions of excitation energies over lines of the same species (see Section~\ref{analytic_exc_sect} below). Figures~\ref{exc_s} and \ref{exc_p} show the correlations between excitation potential and abundance difference (for ionized species) as well as between the difference of ionization and excitation potentials and abundance difference (for neutral species).

\subsection{Analytical estimation of effects} \label{analytic_sect}
As giant stars with exactly plane-parallel atmospheres do not exist in nature, it will be difficult to verify these results with observations. A possible experiment would be to perform an abundance analysis of a sample of binary stars, where one component is a dwarf star close to the zero-age main sequence and the second components are giant stars of varying sizes. When both stars are analysed with a plane-parallel spectrum/atmosphere code, any difference in abundance could be attributed to sphericity effects. However, this requires that all other physical phenomena, like convection, deviation from LTE and scattering, are modelled correctly. Here, in order to conduct a ``sanity check'' on the results, we try to reproduce them using an analytical solution to the problem. This will also help to get a better understanding for the cause of the effects. Such solutions require invoking a few more approximations than used in the numerical model. We follow \citet{Chan:34}, who derived a solution to the equation of radiative transfer (in polar coordinates), including the term containing the dependence of intensity $I$ on the angle of inclination to vertical $\theta$
\begin{equation}\label{transport_equ}
\cos\theta \frac{\partial I}{\partial r} - \frac{\sin\theta}{r} \frac{\partial I}{\partial \theta} = \kappa\rho (S - I),
\end{equation}
where $r$ is the distance from the stellar center (i.e., the radius), $\kappa$ is the opacity, $\rho$ the density and $S$ the source function. In local thermodynamic equilibrium (LTE), $S$ is given by the Planck function $B$. Further, if the mean intensity $J$, the flux $\pi F$, and the radiation pressure integral $K$ are defined as the zeroth, first and second moment of the intensity, respectively, the first Eddington approximation $3K=J$ is used, along with the condition that at the outer boundary of the star $F=2J$ (second Eddington approximation). Radiative equilibrium is expressed by $Fr^2=$constant. The dependence of $B$ on optical depth $\tau$ is then
\begin{equation}\label{B_tau_equ}
B = \frac{1}{2}F_R\left(1 + \frac{3}{2}\int_0^{\tau}\frac{R^2}{r^2}d\tau\right),
\end{equation}
where $\pi F_R$ is the flux at $r=R$, which occurs at $\tau=0$. Note that this definition of $R$ is different from the radii given in Section~\ref{models}. If $\tau_{Ross}=10^{-5}$ is taken as an approximation for $\tau=0$, $R$ is the radius at $\tau_{Ross}=1$ enlarged by the corresponding extension (Fig.~\ref{extension}). We adopt this approach in the following.

The same approximations as above are subsequently used when considering the formation of absorption lines. Now, all radiation quantities are indexed with frequency $\nu$ to indicate that they are defined in the range $\nu$ to $\nu+d\nu$ in continuum regions, and additional primes indicate quantities inside spectral lines.

\subsubsection{The $s\_s$ case}
\citet{Chan:34} assumed a power law variation of $\tau_{\nu}$ with $r$
\begin{equation}\label{tau_equ}
\tau_{\nu} = cr^m,
\end{equation}
where $c$ and $m$ are constants and $m < 0$.
Furthermore, $R$ was assumed to be sufficiently large so that
\begin{equation}\label{R_equ}
   \frac{1}{R^2} \ll \frac{3}{2}\int_0^{\tau_{\nu}}\frac{d\tau_{\nu}}{r^2}.
\end{equation}
Assuming small $|m|$, he derived a surprisingly simple expression for the dependence of normalized line flux on line strength, the latter being measured by the ratio of the coefficients of line to continuous absorption $\eta$,
\begin{equation}\label{s_s_equ}
\frac{F_{\nu}'}{F_{\nu}} = \frac{1}{1+\eta}.
\end{equation}

Equation~\ref{s_s_equ} presumably describes the $s\_s$ case and can be directly compared to a similar expression derived by \citet{Eddi:29} for the plane-parallel case
\begin{equation}\label{p_p_equ}
\frac{F_{\nu}'}{F_{\nu}} = \frac{1+\frac{2}{3}q}{1+\eta+\frac{2}{3}q}.
\end{equation}
Here, $q^2=3(1+\eta)(1+\epsilon\eta)$ includes a parameter $\epsilon$ measuring the importance of collisions (i.e. $\epsilon=1$ in LTE).
Comparing the two expressions, it becomes clear that Eq.~\ref{s_s_equ} by far overestimates the effect of sphericity, since the normalized line flux decreases much more rapidly with line strength than in Eq.~\ref{p_p_equ}, and it approaches zero for large $\eta$, whereas in Eq.~\ref{p_p_equ}, the limiting value is $1/(1+\sqrt{3}/2)=0.54$.
Thus, Eq.~\ref{s_s_equ} cannot be used to estimate the effects for the parameter range studied here.
An inspection of the calculated models shows that neither of the two assumptions used by \citet{Chan:34} are fulfilled. We determine values of $m$ and $c$ by fitting the function given by Eq.~\ref{tau_equ} to the calculated values of $\tau_{5000\AA}$ and $r$ in the range $-5 \le \log\tau_{5000\AA} \le 0$. The ratio of $\frac{1}{R^2}$ and the right hand side of Eq.~\ref{R_equ} is then between 0.5 to 0.7 when using $\tau_{5000\AA}=1$ as upper integration limit and correspondingly larger for smaller upper integration limits.
Furthermore, the values of $|m|$ are rather large (between 100 and 2000).

\subsubsection{The $s\_p$ case}
For the $s\_p$ case we start from a slightly different expression for the normalized line flux, derived for the case where the continuum source function varies linearly with optical depth ($S_{\nu}=B_{\nu}=a+b\tau_{\nu}$):
\begin{equation}\label{s_p_equ}
\frac{F_{\nu}'}{F_{\nu}} = \frac{b+aq}{b+a\sqrt{3}}\cdot\frac{1+\frac{2}{3}\sqrt{3}}{1+\eta+\frac{2}{3}q},
\end{equation}
\citep[see][ his Eq.~22]{Eddi:29}.

\begin{figure}
   \resizebox{\hsize}{!}{\includegraphics[]{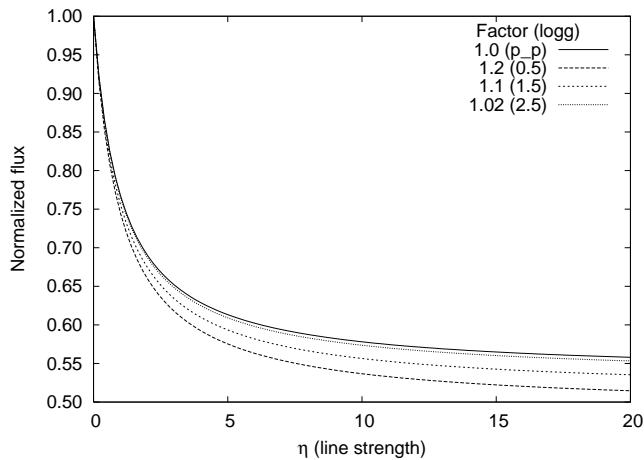}}
      \caption{Normalized flux as a function of line strength showing the $s\_p$ effect as estimated according to Eq.~\ref{s_p_equ}, where $b/a=\frac{3}{8}x c^{2/m} R^2$, for $x=4$ and three different factors $c^{2/m} R^2$ determined from the atmospheric models for different $\log g$ values.
              }
   \label{analytic}
   \end{figure}

The effect of using a spherical model atmosphere compared to a plane-parallel one can then be estimated by comparing corresponding numerical values of the ratio $b/a$.
In LTE, $S_{\nu}=B_{\nu}=C_{\nu}/(e^x-1)$, where $C_{\nu}$ is constant for a given frequency and $x=h\nu/kT$, $T$ being the temperature and $h$ and $k$ the Planck and Boltzmann constants, respectively. The derivative of the logarithm of the Source function with respect to $\tau_{\nu}$ is then
\begin{equation}\label{dlnSdtau_equ}
\frac{1}{S_{\nu}}\frac{dS_{\nu}}{d\tau_{\nu}}=\frac{xe^x}{e^x-1}\frac{1}{T}\frac{dT}{d\tau_{\nu}}.
\end{equation}
For a grey atmosphere, where the continuous absorption is independent of $\nu$, the temperature structure is given by $T^4\propto(1+\frac{3}{2}\tau)$ in the plane-parallel case. Inserting in Eq.~\ref{dlnSdtau_equ} together with the linear source function law, we derive $b/a=\frac{3}{8}x$ at $\tau=0$.
This expression only holds for $x\gtrsim 3$, for which $e^x/(e^x-1)$ can be approximated by 1. Note that the lowest value of $x$ occurring in the atmospheric models studied here is about 3 (for 7200~\AA\ and the hottest models at $\tau_{\rm Ross}=1$).
In the spherical case, using Eq.~\ref{B_tau_equ}, the grey temperature structure can be written as $T^4\propto\left(1+\frac{3}{2}\int_0^{\tau}\frac{R^2}{r^2}d\tau\right)$. We use the same law for $\tau(r)$ as above (Eq.~\ref{tau_equ}) to evaluate the integral in the temperature structure equation and further the derivative in Eq.~\ref{dlnSdtau_equ}. For large values of $|m|$, this leads to $b/a=\frac{3}{8}x c^{2/m} R^2$.
In order to estimate numerical values of the discriminating factor $c^{2/m} R^2$, we resort to determining them from the atmospheric models: for $\log g$ values of 0.5, 1.5 and 2.5, typical values of the factor are 1.2, 1.1, and 1.02, respectively. The resulting variation in normalized line flux is shown in Fig.~\ref{analytic} as a function of line strength $\eta$, for a value of $x=4$ (corresponding to e.g. a wavelength of 6000~\AA\ at a temperature of 6000~K). Also shown is the relation for the $p\_p$ case (Eq.~\ref{p_p_equ}, which derives from Eq.~\ref{s_p_equ} when $x=4$). Comparing this figure to Fig.~\ref{profiles} shows that the analytic estimation of the sphericity effect agrees fairly well with the modelling outcome in this case, at least for lines of species in the minority ionization stage.

The results of this discussion suggest another observational test. When central depths of saturated lines are measured in giant stars of different radii, they are relatively unaffected by star-to-star abundance differences. Those lines having cores that form close to LTE should then be deeper for the more extended stars than for the less extended stars by the amount predicted.

\subsubsection{Dependence on excitation potential} \label{analytic_exc_sect}
The variation of the abundance differences with excitation potential ($\chi$) can under certain circumstances be directly estimated from model temperature differences. We use the assumption that for a weak line of a given species, changes in logarithmic abundance are equivalent to changes in $5040/T\cdot\chi$ for a line from the majority species \citep[see e.g.][ Chapter 14]{gray}. The temperature difference in the region of $\log\tau_{\rm Ross}$ between $-0.1$ and $-0.5$, where most of the lines form, is about 30, 15 and 5~K for models with $\log g = 0.5$, 1.5 and 2.5, respectively. The corresponding variations of abundance difference with $\chi$ are shown in Fig.~\ref{exc_p} (lower plot) as lines and are comparable to the outcome of the detailed models.
For lines of species in the minority ionization stage (all neutral elements except C~I, O~I and S~I, see Table~\ref{species}), the differences will also depend on the ionization potential.

\subsection{Implications for abundance analysis}
The systematically different effects on lines of different ionization stages of the same element imply an effect on ionization equilibrium, which is sometimes used to infer gravities of program stars. Just as an example, we take the star \object{YZ Sgr} from \citet{Luck:04}. At phase 0.249 it has a temperature of 5500~K, $\log g= 1.4$ and $\xi_{\rm t}$ = 3.7~km~s$^{-1}$. The average abundance difference ($p\_p - s\_s$) for Fe~I lines used by \citet{Luck:04}, which are also in our line list, amounts to $+0.03\pm0.01$~dex (for the 5500~K/1.5 model with $\xi_{\rm t}$ = 2 or 5~km~s$^{-1}$). For Fe~II, the difference is $-0.010\pm0.003$~dex. The same difference is obtained by decreasing $\log g$ in the $s\_s$ models by about 0.1~dex or increasing the temperature by about 40~K. These values are equal to the formal uncertainty given for this star.
In the \citet{Take:05} sample, stars at 5000~K have $\log g \ge 2.5$. If we take as an example the star \object{HD 192944} ($\xi_{\rm t}$ = 1.4~km~s$^{-1}$), the abundance differences ($p\_p - s\_s$) are only $+0.009\pm0.006$~dex and $-0.008\pm0.004$~dex for Fe~I and Fe~II, respectively. They can safely be ignored until abundance analyses reach significantly higher internal accuracies compared to today.
Abundance differences for individual lines and/or stellar parameters can be obtained on request from the authors.

\section{Conclusions}

   \begin{enumerate}
      \item We recommend the use of spherical model atmospheres in abundance analyses for $\log g \lesssim 2$ and $4000$~K $\le T_{\rm eff} \le 6500$~K. Thus one can avoid both systematic errors on abundances and differential effects, which can lead to additional uncertainties in stellar parameters. Alternatively, restricting the analysis to weak lines ($W \lesssim 100$~m\AA) and/or majority species will minimize the uncertainties on abundances caused by assuming a plane-parallel geometry.
      \item Geometry has a smaller effect on line formation than on model atmosphere structure.
      It is more important to use a spherical model atmosphere than to be consistent when calculating the spectral lines, i.e. an $s\_p$ calculation is better than $p\_p$ if tools for $s\_s$ calculations are not available.
      Note, however, that this holds only for the range of parameters studied here. For a model of a cool giant with ($T_{\rm eff}$,$\log g$)=(2800~K,-0.6) for example, \citet{Arin:05} showed that an inconsistent treatment of geometry ($s\_p$) results in larger differences in spectral line depth than for a fully plane-parallel calculation.
      \item For both cases ($s\_p$ and $p\_p$), the results presented in Section~\ref{results} are intended to provide a guide to the estimation of systematic errors in abundance analyses due to geometry effects.
      For $s\_p$ calculations, the largest differences encountered with respect to the $s\_s$ case are $-0.1$~dex. In the $p\_p$ case, differences are up to $+0.35$~dex for minority species and at most $-0.04$~dex for majority species.
      The abundance studies of halo stars mentioned in Section~\ref{introduction} contain stars in the $\log g$ range where sphericity effects are important. Although the abundance difference values given in Section~\ref{results} are based on solar abundance models, effects are expected to be similar in low-metallicity models, since the atmospheric extensions are similar.
   \end{enumerate}

\begin{acknowledgements}
We thank all people who have contributed to the development of the current Uppsala version of the MARCS code throughout many years. We thank B. Edvardsson for calculating the grid of MARCS models used for the present work. B. Gustafsson, R.E. Luck, and the referee, C.I. Short, are thanked for several useful comments.
\end{acknowledgements}

\bibliographystyle{aa} 
\bibliography{models}

\begin{thebibliography}{19}
\expandafter\ifx\csname natexlab\endcsname\relax\def\natexlab#1{#1}\fi

\bibitem[{{Andersen}(1999)}]{Ande:99}
{Andersen}, J. 1999, Transactions of the IAU, vol. XXIIIB (Kluwer Academic
  Publishers), p. 141

\bibitem[{{Andrievsky} {et~al.}(2004){Andrievsky}, {Luck}, {Martin}, \& {L{\'
  e}pine}}]{Andr:04}
{Andrievsky}, S.~M., {Luck}, R.~E., {Martin}, P., \& {L{\' e}pine}, J.~R.~D.
  2004, \aap, 413, 159

\bibitem[{{Aringer}(2005)}]{Arin:05}
{Aringer}, B. 2005, in Proc. ESO Workshop: High-resolution IR spectroscopy in
  Astronomy, ed. H.~K\"aufl, R.~Siebenmorgen, \& A.~Moorwood

\bibitem[{{Chandrasekhar}(1934)}]{Chan:34}
{Chandrasekhar}, S. 1934, \mnras, 94, 444

\bibitem[{{Eddington}(1929)}]{Eddi:29}
{Eddington}, A.~S. 1929, \mnras, 89, 620

\bibitem[{{Gray}(1992)}]{gray}
{Gray}, D.~F. 1992, The observation and analysis of stellar photospheres
  (Cambridge University Press)

\bibitem[{{Gustafsson} {et~al.}(2003){Gustafsson}, {Edvardsson}, {Eriksson},
  {Mizuno-Wiedner}, {J{\o}rgensen}, \& {Plez}}]{Gust:03}
{Gustafsson}, B., {Edvardsson}, B., {Eriksson}, K., {et~al.} 2003, in ASP Conf.
  Ser. 288: Stellar Atmosphere Modeling, 331

\bibitem[{{Gustafsson} \& {Olander}(1979)}]{Gust:79}
{Gustafsson}, B. \& {Olander}, N. 1979, \physscr, 20, 570

\bibitem[{{Hauschildt} {et~al.}(1999){Hauschildt}, {Allard}, {Ferguson},
  {Baron}, \& {Alexander}}]{Haus:99}
{Hauschildt}, P.~H., {Allard}, F., {Ferguson}, J., {Baron}, E., \& {Alexander},
  D.~R. 1999, \apj, 525, 871

\bibitem[{{Kupka} {et~al.}(1999){Kupka}, {Piskunov}, {Ryabchikova}, {Stempels},
  \& {Weiss}}]{Kupk:99}
{Kupka}, F., {Piskunov}, N., {Ryabchikova}, T.~A., {Stempels}, H.~C., \&
  {Weiss}, W.~W. 1999, \aaps, 138, 119

\bibitem[{{Luck} \& {Andrievsky}(2004)}]{Luck:04}
{Luck}, R.~E. \& {Andrievsky}, S.~M. 2004, \aj, 128, 343

\bibitem[{{McWilliam} {et~al.}(1995){McWilliam}, {Preston}, {Sneden}, \&
  {Searle}}]{McWi:95}
{McWilliam}, A., {Preston}, G.~W., {Sneden}, C., \& {Searle}, L. 1995, \aj,
  109, 2757

\bibitem[{{Nordlund}(1984)}]{Nord:84}
{Nordlund}, A. 1984, in Methods in Radiative Transfer, ed. W.~{Kalkofen}
  (Cambridge University Press), 211--233

\bibitem[{{Plez}(1990)}]{Plez:90}
{Plez}, B. 1990, Memorie della Societa Astronomica Italiana, 61, 765

\bibitem[{{Plez} {et~al.}(1992){Plez}, {Brett}, \& {Nordlund}}]{Plez:92}
{Plez}, B., {Brett}, J.~M., \& {Nordlund}, A. 1992, \aap, 256, 551

\bibitem[{{Ryan} {et~al.}(1996){Ryan}, {Norris}, \& {Beers}}]{Ryan:96}
{Ryan}, S.~G., {Norris}, J.~E., \& {Beers}, T.~C. 1996, \apj, 471, 254

\bibitem[{{Simmerer} {et~al.}(2004){Simmerer}, {Sneden}, {Cowan}, {Collier},
  {Woolf}, \& {Lawler}}]{Simm:04}
{Simmerer}, J., {Sneden}, C., {Cowan}, J.~J., {et~al.} 2004, \apj, 617, 1091

\bibitem[{{Takeda} {et~al.}(2005){Takeda}, {Sato}, {Kambe}, {Izumiura},
  {Masuda}, \& {Ando}}]{Take:05}
{Takeda}, Y., {Sato}, B., {Kambe}, E., {et~al.} 2005, \pasj, 57, 109

\bibitem[{{Travaglio} {et~al.}(2001){Travaglio}, {Galli}, \&
  {Burkert}}]{Trav:01}
{Travaglio}, C., {Galli}, D., \& {Burkert}, A. 2001, \apj, 547, 217

\end{thebibliography}

\end{document}